\documentstyle[12pt,twoside]{article}
\def\greaterthansquiggle{\raise.3ex\hbox{$>$\kern-.75em\lower1ex\hbox{$\sim$}}}
\def\lessthansquiggle{\raise.3ex\hbox{$<$\kern-.75em\lower1ex\hbox{$\sim$}}}
\newcommand{\beq}{\begin{equation}}
\newcommand{\eeq}{\end{equation}}
\newcommand{\beqa}{\begin{eqnarray}}
\newcommand{\eeqa}{\end{eqnarray}}
\newcommand{\beqan}{\begin{eqnarray*}}
\newcommand{\eeqan}{\end{eqnarray*}}
\newcommand{\ba}{\begin{array}}
\newcommand{\ea}{\end{array}}

\newcommand{\A}{{\cal A}}

\def\nz{\ifmmode {I\hskip -3pt N} \else {\hbox {$I\hskip -3pt N$}}\fi}
\def\zz{\ifmmode {Z\hskip -4.8pt Z} \else
       {\hbox {$Z\hskip -4.8pt Z$}}\fi}
\def\qz{\ifmmode {Q\hskip -5.0pt\vrule height6.0pt depth 0pt
       \hskip 6pt} \else {\hbox
       {$Q\hskip -5.0pt\vrule height6.0pt depth 0pt\hskip 6pt$}}\fi}
\def\rz{\ifmmode {I\hskip -3pt R} \else {\hbox {$I\hskip -3pt R$}}\fi}
\def\cz{\ifmmode {C\hskip -4.8pt\vrule height5.8pt\hskip 6.3pt} \else
       {\hbox {$C\hskip -4.8pt\vrule height5.8pt\hskip 6.3pt$}}\fi}

\def\au{{\setbox0=\hbox{\lower1.36775ex%
\hbox{''}\kern-.05em}\dp0=.36775ex\hskip0pt\box0}}
\def\ao{{}\kern-.10em\hbox{``}}

\normalsize
\begin{document}
\bibliographystyle{plain}

\begin{titlepage}
\begin{flushright}
\today
\end{flushright}
\vspace*{2.2cm}
\begin{center}
{\Large \bf Bose Condensate for Quasifree Fermions }\\[30pt]

Heide Narnhofer  $^\ast $\\ [10pt] {\small\it}
Fakult\"at f\"ur Physik \\ Universit\"at Wien\\

\vfill \vspace{0.4cm}

\begin{abstract} We construct the fluctuation algebra for fermions in a quasifree state and its timedependence for quasifree evolution. We find a Bose-Einstein condensate and study its stability under interaction.

\smallskip
Keywords:  Quasifree evolution, fluctuation algebra, Bose-Einstein-condensate, dynamical stability
\\
\hspace{1.9cm}

\end{abstract}
\end{center}

\vfill {\footnotesize}

$^\ast$ {E--mail address: heide.narnhofer@
univie.ac.at}
\end{titlepage}
\section{Introduction}
Product states and quasifree states are the examples where we have complete control over all expectation values of every quasilocal operator. Every such expectation value can be constructed out of a few constituents. That is the reason that already many years ago H. Araki said, that further relevant insights on these systems are out of reach and research in this direction can only have the purpose to keep one busy. As being retired this however was exactly my motivation. Nevertheless the result in this enterprise came a bit as a surprise to me, in its simplicity, but also because it is related to stability properties in thermodynamics.

\cite{HGV} suggested the construction of a fluctuation algebra as a state dependent algebra on macroscopic size but still with quantum features. With appropriate scaling it turned out to be a CCR algebra. This algebra is not only a mathematical construction but in fact was used in \cite{P} to observe quantum effects as entanglement on macroscopic size. The result was interpreted in terms of the fluctuation algebra of a product state in \cite{NT}. Also macroscopic quantum effects of superconductivity can be related to this algebra (\cite{INT}, \cite{BCFN}). On the mathematical level \cite{M} improved the considerations showing that in equilibrium states all necessary convergence properties are satisfied and that in addition time evolution defined as automorphism group on the fluctuation algebra again satisfies the KMS-condition. This result has consequences for the commutation relations, because not every quasifree evolution of bosons allows the existence of KMS-states. In addition the question arises whether on these fluctuation algebras Bose-Einstein-condensation can be observed.

In this note we will first repeat shortly the relevant properties of a CCR-algebra, especially in its relation to Bose-Einstein-condensation. In chapter 3 we repeat the definition of the fluctuation algebra. In chapter 4 we will study as a comparison product states and show the existence of a condensate. The time evolution is quasiperiodic and the condensate has no relation to a phase transition.

In chapter 5 we will turn to quasifree states. Again we can construct the condensate as the fluctuations corresponding to quadratic local operators as an abelian algebra, and this holds for all quasifree states.  It is pointwise time invariant.  We concentrate on the fluctuation algebra corresponding to other gauge invariant quasilocal elements. The time evolution on this algebra is quasifree and inherits asymptotic abelianess from the quasilocal level. The corresponding spectrum is positive if on the quasilocal level the state satisfies the KMS-condition. In general it can contain positive and negative parts. Introducing interaction on the quasilocal level with the appropriate scaling (that however has a negligible effect on the quasilocal level) it corresponds to an inner perturbation on the fluctuation level. Convergence in time of this perturbed evolution is controlled by scattering theory and convergence is only guaranteed if the initial state satisfies the KMS-condition.

\section{The CCR-algebra}
We follow the description in \cite{BR}. The CCR-algebra is built by Weyloperators corresponding to a map $f \in H$, $f$ elements of a real linear space with a nondegenerate symplectic bilinear form $\sigma (f,g)= -\sigma (g,f)$. Except if $H$ has odd finite dimension there exists an operator $J$ on $H$ with $J^2=-1$ and
$$\sigma (Jf,g)=-\sigma (f,Jg).$$
This operator allows to extend $H$ to a complex Hilbert-space by defining $$(\lambda _1 +i\lambda _2)f=\lambda _1 f+\lambda _2 Jf$$
so that on this Hilbert-space the scalar product is defined as
$$\langle f|g\rangle =\sigma (f,Jg)+i\sigma (f,g).$$
Eigenvectors in this complex Hilbertspace correspond to functions $$\chi _i,\eta _j$$ satisfying
$$\sigma (\chi _i,\chi _j)=\sigma (\eta _i, \eta _j)=0, \quad \sigma (\chi _i, \eta _j)=\delta _{i,j}$$

so that
$$\langle\chi _i|\eta _j\rangle =0, \quad \langle \chi _i|\chi _j\rangle =\langle \eta _i|\eta _j\rangle =\delta _{i,j}$$
The Weyloperators satisfy $$W(f)^*=W(-f), \quad W(f)W(g)=e^{-i\sigma (f,g)/2}W(f+g)$$
If there exists a real invertible operator $T$ on $H$ such that $$\sigma (Tf,Tg)=\sigma (f,g)$$
then it defines an automorphism $\gamma $ on the Weyl-algebra via $$\gamma W(f)=W(Tf)$$
Quasifree states on the Weyloperator are defined by an operator $A$ over the Hilbertspace
$$\omega (W(f))=e^{-\langle f|A|f\rangle } \quad 0<A<1/4$$
They are KMS-states with respect to the time evolution
$$\tau _t W(f)=W(e^{iht}f)$$
with respect to the temperature $1/\beta $ and the chemical potential $z<1$ for
$$A=\frac{(1+ze^{-\beta h})(1-ze^{-\beta h})}{4}.$$
If the symplectic form is degenerate, i. e. if for some $f_0$
$$\sigma (f_0,g)=0 \quad\forall g$$
then the expectation value $\omega (W(f_0))$ can be chosen freely, i.e. we can interpret that $f_0$ corresponds to a groundstate and can be occupied by an arbitrary number of Bosons. Therefore we can talk of a Bose-condensate for the mode $f_0.$

For free bosonic systems we consider $h_{\Lambda }>0$ over a finite region $\Lambda $ but approaching $\geq0$ as lower limit for $\Lambda $ tending to infinity. Therefore in the thermodynamic limit we can obtain different occupations of the groundstate-mode corresponding to $p=0$ depending how we let approach
$$\lim _{\Lambda \rightarrow \infty }z_{\Lambda }=1.$$
 The essential feature is the fact that $$\omega (W(\tau _t f_0 )=\omega (W(f_0)).$$
 The representation of the state contains a nontrivial center. That we talk about Bose-condensation as a special phase transition corresponds to the fact that differently to other phases the symmetry that has to be broken to obtain extremal KMS-states is now the gauge-symmetry and therefore only a mathematical but not a physical possibility.

\section{The Fluctuation algebra}
It was introduced in \cite{HGV}. Assuming $\omega $ is a space translation invariant state, exponentially clustering in space, we can define in the GNS-representation for every quasilocal operator $A=A^*$ and $\sigma _x$ space-translation the fluctuation-operator
$$F(A)=\lim_{\Lambda \rightarrow \infty }\Pi _{\omega }exp (\frac{i}{\sqrt{|\Lambda |}}\int _{\Lambda }dx (\sigma _x A-\omega (A))$$
This limit exists (\cite{HGV},\cite{M}) and satisfies, suppressing the $\omega $ dependence
\beq F(A)F(B)= e^{i\sigma (A,B)}F(B)F(A)\eeq
where
\beq \sigma (A,B)=\lim _{\Lambda }i\int _{\Lambda }dx \omega ([A, \sigma _x B]) \eeq
is the desired symplectic form necessary for the definition of the CCR-algebra. $\omega $ defines a quasifree state on this algebra
\beq \bar{\omega }(F(\alpha A))=e^{-\alpha ^2 w(A)} \eeq
which by expanding in $\alpha $ can be seen to satisfy
\beq w(A) = \lim _{\Lambda } \int _{\Lambda }dy (\omega (A\sigma _y A)-\omega (A)^2)\geq 0 \eeq
Assume $\tau _t $ is a time evolution on the quasilocal algebra, locally analytic, then it defines a time evolution on the fluctuation algebra satisfying
\beq \bar{\tau _t}F(A)=F(\tau _t A) .\eeq
If $ \omega $ is a KMS-state for the quasilocal algebra then $\bar{\omega }$ inherits the analyticity properties and is a KMS-state for the fluctuation algebra with respect to $\bar{\tau _t}.$

\section{The Fluctuation-algebra for product states}
This is the example which worked in \cite{P} with the restriction to a finite subalgebra of the fluctuation algebra. Here we construct it in more generality so that we can compare it with the fluctuation algebra of quasifree fermionic states. We start with the quasilocal algebra on the chain $\A = \otimes _x \sigma _x M_n$ with $M_n$ a n-dimensional matrix algebra and $\sigma _x $ the automorphismgroup of space translations. The time evolution acts strictly locally and is translation invariant
\beq \tau _t\Pi \otimes m_x=\Pi \otimes e^{it\sigma _x h}m_x e^{-it\sigma _x h} \eeq
and
\beq \omega (\Pi \otimes m_x)=\Pi \omega (m_x) =\Pi \omega (\otimes e^{it\sigma _x h}m_x e^{-it\sigma _x h})\eeq
The symplectic form for the fluctuation algebra reads
\beq \sigma (\Pi \otimes m_x, \Pi \otimes n_y)=\Pi i\omega ([m_x,n_x]) \eeq
Let $h=\sum _j h_j e_{jj}$ with matrix units $e_{ij}$. Then
\beq \sigma (\Pi _k \otimes h_{j_k}e_{j_k,j_k}, \Pi \otimes n_x) =0\eeq
Our symplectic form is degenerate and $F(\Pi _k \otimes h_{j_k}e_{j_k,j_k})) $ forms an abelian $\bar{\tau _t}$ invariant algebra that we can call the Bose condensate of the fluctuation algebra. Except for the pure state it consists of several modes, and for a time-invariant pure state, i.e.$\rho _0=e_{jj}$ also the center is pure. In fact it is the maximal Bose condensate . For every other element $F(\sum\Pi m_x)$ we can find another $F$ so that they do not commute: we can write every element in the form
$ A =\sum (c_{jk}e_{jk}\otimes B_{jk}+\bar{c}_{jk}e_{kj}\otimes B_{jk}^*)$ and choose $C=(ce_{jk}+\bar{c}e_{kj})\otimes 1$ and adjust $c$ so that the expectation value of the commutator does not vanish.

The CCR-algebra splits into subalgebras of different range. They have dimension $n^2k^2-nk=nk(nk-1)$ and therefore satisfy the demand for the construction of $J$. Therefore we obtain a sum over Hilbertspaces $H_{\alpha}$ where $\chi _{\alpha }$ can be interpreted as $p_{\alpha }$ and $\eta _{\alpha }$ as $x_{\alpha }$ with $[x_{\alpha }, p_{\alpha }]=i.$

\section{The Fluctuation-algebra for quasifree states}
We consider the Fermi-algebra over a Hilbert-space $H(p)\otimes H(\alpha )$ built by creation and annihilation operators $a(f(p,\alpha )), a^*(f(p,\alpha ))$ satisfying
\beq [a(f(p,\alpha )),a(g(p, \alpha ))]_+=0 \quad [a(g(p,\alpha )),a^*(f(p,\alpha ))]=\int \bar{g}(p,\alpha )f(p,\alpha )dpd\mu(\alpha ) .\eeq
We are especially interested in $H(\alpha )=1$ but want to keep the possibility that $H(\alpha ) $ is finite dimensional and represents the spin.
A quasifree state is given by
\beq \omega (a^*(f_1)..a^*(f_n)a(g_m)..a(g_1))=\sum _{\Pi}(-1)^{\Pi}\delta _{nm}\Pi \omega (a^*(f_i)a(g_{\Pi _i}))\eeq
where the sum runs over all permutations. Therefore the state is determined by the two-point-function that reads
\beq \omega (a^*(f)a(g))=\langle g|\rho |f\rangle = \int \bar{g}(p,\alpha ) f(p, \alpha )\rho (p,\alpha )dpd\mu (\alpha  )\eeq
with $0\leq \rho (p, \alpha )\leq 1.$

We concentrate on gauge invariant elements $\Pi _{i=1}^na^*(f_i)a(g_i)$. The commutator of an operator of order two with gauge invariant elements is given by
\beq [a^*(f_1)a(f_2), a^*(g_1)..a^*(g_n)a(h_n)..a(h_1)]_-=\sum (-1)^j\langle f_2|g_j\rangle a^*(g_1)..a^*(g_j)^{elim}a^*(g_n)a(h_n)..a(h_1) \eeq
$$+\sum (-1)^j \langle h_j|f_1\rangle a^*(g_1)..a^*(g_n)a(h_n)..a(h_j)^{elim}..a(h_1) $$
where we indicated the eliminated contribution. Its expectation value reads
\beq \sum (-1)^j(-1)^{\Pi}\langle f_2|g_j\rangle \Pi \omega (a^*(g_k)a(h_{\Pi (k)})
+\sum (-1)^j(-1)^{\Pi' }\langle h_j|f_1\rangle \Pi\omega (a^*(g_k)a(h_{\Pi (k)})\eeq
where we identify for the permutation $f_1 =g_0$ and $f_2 =h_0.$
For the fluctuation algebra we have to consider selfadjoint elements, therefore we have to include the terms with $f_1$ replaced by $f_2$ and $g_j$ replaced by $h_j.$ Especially we can collect
\beq (\langle g_j|f_i\rangle \omega (a^*(g_j)a(f_i))-\langle f_i|g_j\rangle \omega (a^*(f_i)a(g_j)))\omega (\Pi _{\neq i\neq j} a^*..a)\eeq
which taking into account that $f_j$ is shifted and the rest is kept fixed and contributes as
\beq\int dx \int \bar{g} _j f_i e^{ipx}dp \int g_j\bar{f}_i g_j e^{-iqx}\rho dq= \int \bar{g}_jg_j\bar{f}_if_i\rho dp \eeq
and the two terms cancel one another. It follows that $F(a^*(f)a(g)+a^*(g)a(f))$ belongs to the center. In general it is not trivial. Especially
\beq \int dx  (\omega (a^*(f)a(f)a^*(f_x)a(f_x))-\omega (a^*(f)a(f))^2)=\int \bar{f}^2f^2\rho (1-\rho)dp\eeq
This only vanishes if the state is pure.
According to \cite{M} elements of the center have to be time-invariant, if the state satisfies the KMS-condition. But in fact this is not even necessary. Time invariance holds if
\beq F(a^*(f)a(g)+a^*(g)a(f))F(-a^*(f_t)a(g_t)-a^*(g_t)a(f_t))=1 \eeq
or according to (4) if $$w(a^*(f)a(g)+a^*(g)a(f)-a^*(f_t)a(g_t)+a^*(g_t)a(f_t))=0.$$ Most of the contributions are obviously time invariant. Less evident it is for
\beq \int dx \omega (a^*(f)a(g_{tx}))\omega (a^*(f_{tx})a(g))=\int dx \int f\bar{g}e^{-i(px+h(p)t)}\rho dp \int f \bar{g}e^{i(qx+ih(q)t)}\rho dq \eeq
$$=\int f^2\bar{g}^2\rho^2dp.$$
Again this expression is time-invariant. For $t=0$ (18) reduces to $0$ and being time-invariant therefore also for all $t$. Notice, that again it was essentiell, that the $x$-integration only happens for the scalar product of a twopoint function with a quadratic expectation value or two such expectation-values. If we consider the fluctuation operators of higher products then the $x$-integration produces a $\delta (\sum p_j -\sum q_k)$ that does not implement cancelation for the $t$-dependent terms. However being interested in the long time behaviour of the fluctuation algebra all relevant terms are of the form
\beq \int dp_1..dq_k \bar{f}_1g_1...\bar{f}_kg_k e^{it(h(p_1)..-h(q_k))} \delta (\sum p_j-\sum q_l)\eeq
so that the time-evolution is neither trivial nor periodic but by Riemann Lebesque becomes asymptotically abelian if it is so on the quasilocal level.

Taking into account that the Hilbert-space is enlarged by $H(\alpha )$ we can concentrate on gauge invariant states satisfying
\beq \omega (a^*(f_{\alpha })a(g_{\beta }))=0 \quad \alpha \neq \beta \eeq
As a consequence the center is enlarged by the elements $F(a^*(f_{\alpha })a(g_{\beta })+h.c.).$
Either
\beq \omega (a^*(f_{\alpha })a(g_{\alpha }))\neq \omega (a^*(f_{\beta })a(g_{\beta }))\eeq
and the considerations remain unchanged. If on the $H(\alpha )$ we have degeneracy, this corresponds to the fact, that on this part of the algebra we have the trace, that let the expectation value of commutators vanish and also makes it necessary that $h_{\alpha }(p)=h_{\beta }(p)$. Again the center is built by the gauge invariant operators of order two.

We are interested whether the state on the fluctuation algebra can give us some information on stability properties of the underlying quasilocal algebra, i.e. we want to disturb the quasifree time evolution by an additional interaction, that however is of the order $|\Lambda |^{-1/2}$, therefore negligible on the quasilocal level but felt on the level of the fluctuation algebra.

Let the perturbation correspond on the quasilocal level to a selfadjoint operator $G$. On the level of the fluctuation algebra the unperturbed time evolution is implemented by $\bar{H}$ and corresponding to $G$ perturbed by $b(\bar{g})+b^*(\bar{g})$, where $b,b^*$ are bosonic annihilation and creation operators and $\bar{g}$ the function in the Hilbertspace underlying the CCR-algebra. The perturbed time evolution solves
\beq \frac{d}{dt}b(\bar{f_t})= i[\bar{H},b(\bar{f_t})]+i[b^*(\bar{g}), b(\bar{f_t})]\eeq
which is satisfied by
\beq b(\bar{f_t})=b(exp(i\bar{H}t+i\langle \bar{g}|\int e^{it'\bar{H}}dt'\bar{f}\rangle \eeq
We are interested whether the perturbed evolution converges for $t\rightarrow \infty .$ This happens, when scattering theory applies, i. e. when
\beq \lim _{T\rightarrow \infty }\int _0^T dt \langle \bar{g}|e^{i\bar{H}t}\bar{f}\rangle = \lim _{
T\rightarrow \infty } \int _0^T dt \int dx\omega ([G,\tau _t \sigma _x F]) \eeq
exists. We have already observed, that the time evolution of the fluctuation algebra inherits asymptotic abelianess from the quasilocal level. But now we can inherit strong convergence of the scattering operator for the fluctuation algebra from the rate, how fast commutators for the quasilocal algebra converge in norm. As in (20) convergence is determined after $x$-integration by
\beq e^{it(\sum h(p_j)-\sum h(q_k))}\delta (\sum p_j -\sum q_k)\eeq
i.e. by an exponent that can vanish where the term that is integrated does not vanish. Therefore applying van der Carput or the method of stationary phases \cite{MW},\cite{SM} Riemann-Lebesgue only guarantees decay up to order $t^{-1/k}$ and the time integration does not converge.

There is however the possibility that by good coincidence the integrand vanishes together with the exponent, and this happens, when our initial quasifree state and therefore also the state over the fluctuation algebra satisfies the KMS condition. Then we deal with
\beq \int dt e^{i\Phi (p)t}(1-e^{-\beta \Phi (p)})\chi(p)d\mu (p) \eeq
in the Hilbert-space corresponding to the CCR-algebra and we can use that for $\Phi (p)=0$ also $1-e^{-\beta \Phi (p)}=0$ and we can use the theorem of van der Carput to obtain integrability in $t.$

This demonstrates that equilibrium states are more stable with respect to perturbation of the dynamics as it was shown in \cite{PW}, \cite{HKTP}, there for local perturbations and now for global but weak perturbations on the fluctuation level.

\section{Conclusion}
We have examined properties of the fluctuation algebra of quasifree states in comparison to the one of product states. In the later case time invariant operators create the center of the fluctuation algebra, trivial only when we start with the tracial state. For quasifree states with a time-evolution that is asymptotically abelian no time invariant operators on the local level are available, however for the fluctuations corresponding to gauge invariant elements of quadratic type create the center, that is pointwise invariant under time evolution. The same abelian subalgebra of the fluctuation algebra exists for states that are not quasifree being constructed out of the two point function, however we loose time invariance and also commutativity with other fluctuation operators.
In addition we studied perturbation of the dynamics of the fluctuation algebra, corresponding to perturbation of the dynamics by interaction on the quasilocal level, that however are too small to be observable on the local level. Nevertheless they become observable on the fluctuation level. Starting with a time-invariant state on the quasilocal level we can construct a perturbed time dependent state on the fluctuation level that however does not converge in general to an invariant state.  But for equilibrium states we can use the perturbation theory in the imaginary time direction \cite{BR} and obtain a time invariant perturbed state. The center of the fluctuation algebra represents a stability region with respect to perturbation of the dynamics when the rest  of the fluctuation algebra is not stable any more. The same abelian algebra can be constructed for states that are not quasifree, as the construction is based only on the twopoint-function. However, neither it stays invariant under the time evolution nor do its elements commute with the rest of the fluctuation algebra. It would be interesting whether also for invariant states of interacting systems another subalgebra survives as center of the fluctuation algebra.

\bibliographystyle{plain}

\end{document}